\documentclass[aip,rsi,amsmath,amssymb,graphicx,preprint]{revtex4-1}

\usepackage{amsmath}
\usepackage{graphicx}
\usepackage{amssymb}
\usepackage{dcolumn}
\usepackage{bm}
\usepackage{color}
\usepackage{float}
\usepackage{upgreek}
\usepackage[final]{hyperref}
\hypersetup{
	colorlinks=true,    
	linkcolor=blue,     
	citecolor=blue,     
	filecolor=magenta,  
	urlcolor=blue         
}
\usepackage[normalem]{ulem}

\makeatletter

\begin{document}
\title{Diffusion of interacting particles in a channel with reflection boundary conditions}

\author{Narender Khatri}\thanks{narenderkhatri8@iitkgp.ac.in}
\affiliation{Department of Physics, Indian Institute of Technology Kharagpur, Kharagpur-721302, India}
	
\author{P. S. Burada}\thanks{Corresponding author: psburada@phy.iitkgp.ac.in}
\affiliation{Department of Physics, Indian Institute of Technology Kharagpur, Kharagpur-721302, India}
\affiliation{Center for theoretical studies, Indian Institute of Technology Kharagpur, Kharagpur-721302, India}

\date{\today}
	
\begin{abstract}  

The diffusive transport of biased Brownian particles in a two-dimensional symmetric channel is investigated numerically considering both the no-flow and the reflection boundary conditions at the channel boundaries. Here, the geometrical confinement leads to entropic barriers which effectively control the transport properties of the particles. 
We show that compared to no-flow boundary conditions, the transport properties exhibit distinct features in a channel with reflection boundary conditions. 
For example, the nonlinear mobility exhibits a nonmonotonic behavior 
as a function of the scaling parameter $f$, which is a ratio of the work done to the particles to available thermal energy. 
Also, the effective diffusion exhibits a rapidly increasing behavior at higher $f$. 
The nature of reflection, i.e., elastic or inelastic, also influences the transport properties firmly. 
We find that inelastic reflections increase both the mobility and 
the effective diffusion for smaller $f$. 
In addition, by including the short range interaction force between the Brownian particles, the mobility decreases and the effective diffusion increases for various values of $f$. These findings, which are a signature of the entropic nature of the system, can be useful to understand the transport of small particles or molecules in systems such as microfluidic channels, membrane pores, and molecular sieves.
\end{abstract}

\maketitle

\section{Introduction}

Diffusion is a ubiquitous feature controlling the dynamics of many physical, chemical, and biochemical processes. \cite{Karger} A deep understanding of this process is required to effectively control the mass and charge transport of small objects whose size ranges from microscale to nanoscale. \cite{Karger, Hille} The diffusion of molecules or small particles in confined environments has been an intense area of research in the last few decades due to its major role in processes such as catalysis, osmosis, particles selectivity, and particles separation. \cite{Karger,Siwy,Keyser} The geometrical confinement controls the volume of the phase space accessible to the particles due to which, the entropic barriers arise and influence the transport properties of the Brownian particles in these systems. \cite{Jacobs,Zwanzig,Burada_prl,Burada_phd,Marten,Pineda,Ai_jcp,Das_jcp,
Hanggi_PNAS,Ghosh_1,Ghosh_2, Ghosh_3}  

In one-dimensional energetic landscapes, the transport properties are controlled by the applied force, the height of the potential barrier, and the noise intensity.\cite{Hanggi_rev,Costantini,Hanggi_prl,Lindner}
For example, the nonlinear mobility increases with the noise strength. \cite{Kramers} 
Here, noise assists the particle to overcome the potential barrier, which is independent of the noise strength, whereas in higher dimensional confined systems, an opposite behavior 
can be observed.\cite{Burada_prl,Burada_phd,Bhide} 
In these systems, like aforementioned, the entropic barriers solely control the diffusion process. In this case, the height of the entropic barriers is a function of the noise strength. More noise means less mobility and high diffusion. This is a clear signature of the entropic barriers. The entropic effects are ubiquitous in systems such as biological cells, ion channels, nanoporous materials, zeolites, microfluidic devices, ratchets, and artificial channels. \cite{Karger,Hille, Reza, Han,Brangwynne,Freyhardt,Davis_nature1,Marchesoni_rev, Burada_prl2,Li_pre, Kettner,Matthias}

Earlier, to understand the diffusion of noninteracting Brownian particles, in the absence of advection effects, in confined periodic systems, Zwanzig proposed the modified Fick-Jacobs equation \cite{Zwanzig} with a position dependent diffusion coefficient $D(x)$. Later, several other researchers proposed various forms of $D(x)$ to further improve the accuracy of the Fick-Jacobs equation. \cite{DR_pre,Kali1_pre,Marten} 
To understand the diffusive behavior of noninteracting Brownian particles, in the presence of advection effects, in confined geometries, H\"anggi and co-workers used the same Fick-Jacobs equation. \cite{Burada_prl,Burada_phd,Burada_cpc,Burada_pre}
They have analytically calculated the transport properties of the Brownian particles using the mean first passage time (MFPT) approach. It has been observed that the transport properties can be effectively controlled by a single scaling parameter $f := FL/k_BT$, which is the ratio of work done to the particle and the available thermal energy.
Also, they have reported that the shape of the channel geometry plays a vital role in the diffusive behavior of the particles. Later, several researchers investigated the same problem in detail under various conditions. \cite{Ai_pre, Ai,Borro,PK_jcp,Bauer,Martin,Wang} Note that in most of the studies, no-flow boundary conditions have been used in the numerical simulations for simplicity.
Under no-flow boundary conditions, if the new desired position of the particle is outside the channel, then the Gaussian random number associated with the thermal noise is changed until the new position of the particle is found to be inside the channel. 
In some other studies, people consider that if the new position is outside the channel, then it is discarded and stuck to the old position. \cite{Ai}
However, the analytical description is independent of the boundary conditions. In general, the nature of the boundary conditions can play a pivotal role in the diffusive behavior of the Brownian particles in these confined environments. For instance, when the particle strikes the channel wall and if it gets reflected (reflection boundary conditions), the resident time of the particle in a given cell of the channel (see Fig.~\ref{fig:chw}) may increase. This may influence the transport properties of the particles. In addition, the energy loss due to collisions with the channel walls and the interaction between the particles may further dictate the transport characteristics of the particles. 

In this paper, we study the transport of interacting Brownian particles in a symmetric confined channel with both elastic and inelastic reflection boundary conditions. 
The short range interaction force between the Brownian particles is calculated using the Lubrication theory. \cite{Cooley}
The Lubrication theory provides the hydrodynamics effect on a particle due to its neighbors. Therefore, the interaction force is given by the action of the fluid stress, and it plays a significant role when the particles are confined.
We focus on finding how the reflection boundary conditions, the interaction between the particles, and the aspect ratio of the channel influence the diffusive behavior of the Brownian particles in this confined environment. 

The rest of the article is organized as follows: 
In Sec.~\ref{sec:model}, we introduce our model for the Brownian particles in a 
two-dimensional symmetric channel. 
In Sec.~\ref{sec:3}, we study the diffusive behavior of noninteracting particles in a symmetric channel with elastic reflection boundary conditions. 
The impact of inelastic reflection boundary conditions is discussed in Sec.~\ref{sec:4}. The effect of interaction between the particles is investigated in Sec.~\ref{sec:5}. 
Finally, we present our main conclusions in Sec.~\ref{Conclusions}.

\section{Model} 
\label{sec:model}

\begin{figure}[htb]
\centering
\includegraphics[scale = 1.0]{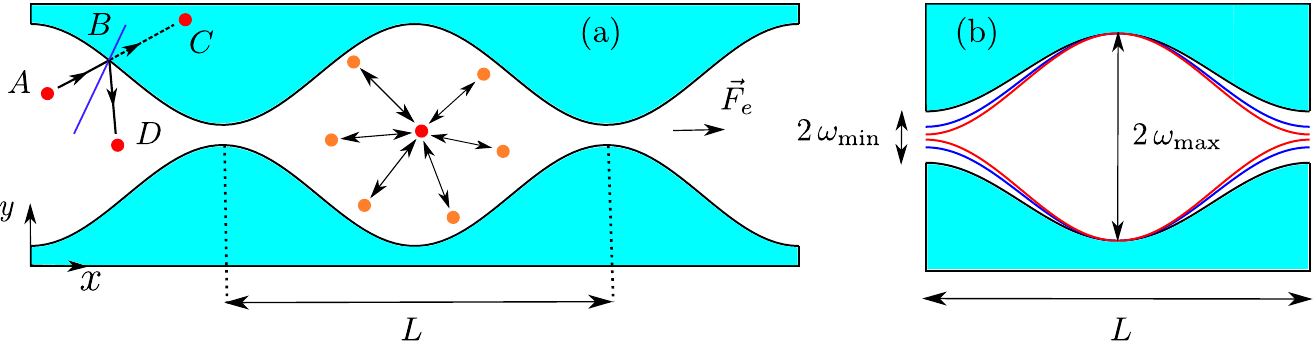}   	
\caption{(Color online) Schematic illustration of a symmetric channel, with the periodicity $L$, confining the motion of Brownian particles which are subjected to a constant external force $\vec{F_e}$ and considering the 
short range interaction force between the particles is depicted in (a). 
The reflection boundary conditions at the channel walls assure the confinement of particles inside the channel. 
(b) Sketch of a single cell of the channel with different aspect ratio $\epsilon = \omega_\mathrm{min}/\omega_\mathrm{max}$.}
\label{fig:chw}
\end{figure}

Consider the motion of a Brownian particle, suspended in a two-dimensional symmetric channel, driven by a constant external force $\vec{F_e}$, along the direction of the channel, and the interaction force $\vec{F}_\mathrm{int}$ due to its neighboring particles (see Fig.~\ref{fig:chw}). 
Here, we assume that the viscous force from the surrounding heat bath dominates over the inertial force of the particle. In this over-damped regime, \cite{Purcell} the equation of motion of a Brownian particle is given by the Langevin equation, 
\begin{equation}
  \eta \frac{d \vec{r}}{d t} = \vec{F}_{e} + \vec{F}_\mathrm{int} + \sqrt{\eta k_B T} \vec{\xi}(t),
\label{eq:Langevin1}  
\end{equation}
where $\vec{r}$ is the position vector of the particle in two dimensions, 
$\eta$ is the friction coefficient, 
$k_B$ is the Boltzmann constant, and 
$T$ is the temperature of the heat bath. 
The thermal fluctuations due to the coupling of the particle with the surrounding heat bath are modeled by a zero-mean Gaussian white noise $\vec{\xi}(t)$, obeying the 
fluctuation-dissipation relation 
$\langle \xi_i (t)  \xi_j (t') \rangle = 2 \delta_{ij} \delta(t-t')$ for $i, j = x, y$.    

The short range interaction force on a particle $i$ due to its neighbors is calculated using the Lubrication theory. \cite{Cooley} 
It reads
\begin{equation}\label{Lubrication}
\displaystyle{\vec{F}_\mathrm{int} =  \sum_{j = 1, ~j \neq i}^{n} \frac{\sigma_{ij}}{d_{ij}}\left( \cos\theta_{ij}\, \hat{x} + \sin\theta_{ij} \, \hat{y} \right)}, 
\end{equation} 
where the sum is taken over all its nearest neighbors, 
$\sigma_{ij}$ is the interaction strength between the particles $i$ and $j$, 
$d_{ij}$ is the corresponding distance between the particles, 
$\theta_{ij}$ is the angle that $d_{ij}$ makes with the channel axis ($x-$ axis), 
$\hat{x}$ and $\hat{y}$ are the unit vectors along the $x$ and $y$ directions, respectively. 
Note that $\sigma_{ij} > 0$ (or $\sigma_{ij} < 0$) means the particles are repelling 
(or attracting) each other. 
In the limit of high density of particles, this interaction force can be approximated as 
\begin{equation}
\vec{F}_\mathrm{int} = k_\mathrm{int} \, ( \cos \theta \,\hat{x} + \sin \theta \,\hat{y}),
\label{eq:Vicsek}
\end{equation}
where $k_\mathrm{int}$ denotes the total strength of the interaction force on a particle due to its neighbors and $\theta$ is a random variable that can 
have values between $0$ to $2 \pi$. 
In this limit, Eq.~(\ref{eq:Vicsek}) is similar to the Vicsek interaction \cite{Vicsek} which has been used extensively to study the collective 
behavior of active particles. 
   
The shape of a two-dimensional symmetric and spatially periodic channel is described by its half-width (see Fig.~\ref{fig:chw}),
\begin{equation} \label{wallf}
\omega(x)= a \sin\left(\frac{2\pi x}{L}\right) + b,  
\end{equation}
where $L$ is the periodicity, and the parameters $a$ and $b$ control the slope and width of the channel. 
Note that to allow the particles from one cell to the other, the condition $b > a$ should be satisfied. 
The half-width at the bottleneck of the channel is given by $\omega_\mathrm{min} = b - a$. 
The maximum half-width of the channel is given by $\omega_\mathrm{max} = b + a$. 
The ratio of these two widths defines the dimensionless aspect ratio given by 
\begin{eqnarray}
\epsilon = \frac{\omega_\mathrm{min}}{\omega_\mathrm{max}};~~~~~~ 0< \epsilon \leqslant 1. 
\end{eqnarray}
Different symmetric channels are identified by different values of the aspect ratio 
$\epsilon$, e.g., $\epsilon = 1$ corresponds to a flat channel.
In terms of $\epsilon$, the relation between the parameters $a$ and $b$ is given by 
$b = a (1 + \epsilon)/(1 - \epsilon)$. 
 
For the sake of a dimensionless description, we rescale the length variables by the periodicity of the channel $L$ and the time variables by $\tau = \eta L^2/(k_B T)$, 
which is the characteristic diffusion time. \cite{Burada_prl} 
In dimensionless variables, the Langevin equation reads 
\begin{equation}\label{Langevin2}
\frac{d \vec{r}}{d t} = \vec{f} +  \vec{f}_\mathrm{int} + \vec{\xi}(t),
\end{equation}
where the dimensionless external force becomes 
$\vec{f} = f \hat{x}$ with $f = F_e L / (k_B T)$, 
which is the ratio of work done to the particle due to the external force and the available thermal energy. 
The dimensionless interaction force becomes 
$\vec{f}_\mathrm{int} = \vec{F}_\mathrm{int} L/(k_B T) = k ( \cos \theta \,\hat{x} + \sin \theta \,\hat{y})$, 
where the scaled interaction strength is defined as $k = k_\mathrm{int} L/(k_B T)$.
Note that in the case of purely energetic systems, e.g., one-dimensional systems, 
the driving forces, $\vec{F}_e$ and $\vec{F}_\mathrm{int}$, and the temperature $T$ are independent variables, 
whereas in systems of entropic nature, these quantities are coupled and characterize the dynamics of systems. \cite{Burada_phd}

In contrast to earlier studies, while solving the Langevin equation $(\ref{Langevin2})$ numerically, in the current study, we have considered that the particle reflects at the channel boundary (elastically or inelastically) to ensure the confinement within the channel. 
Figure~\ref{fig:chw}(a) shows the reflection of a particle at the channel boundary. 
Let us say, the initial position of the particle was $A(x_1, y_1)$ and in the next instant of time, the position of particle is at $C(x_2, y_2)$, which is outside of 
the channel boundary [see Fig.~\ref{fig:chw}(a)]. 
The line joining the points $A$ and $C$ intersects the channel boundary at a point $B(p, q)$, i.e., the reflection point which can be calculated numerically using the bisection method.
Then, the incident angle between $AC$ with the normal at the reflection point $B$ is given by $\theta_i = \tan^{-1} (m_2) - \tan^{-1} (m_1)$,  where $m_1$ and $m_2$ denote the slopes of lines $AC$ and normal, respectively. 
The corresponding reflection angle is $\theta_r = \tan^{-1}(\tan\theta_i/e)$, where $e$ is the coefficient of restitution which characterizes the type of reflection, i.e., elastic or inelastic.
For example, $e = 1$ corresponds to the elastic reflection and $e = 0$ corresponds to the perfect inelastic reflection. 
The momentum conservation and the energy balance lead to the relation $(y_3 - q)^2 + (x_3 - p)^2 = l^2 ( \sin^2 \theta_i + e^2 \cos^2 \theta_i),$ 
where $ l^2 = (y_2 - q)^2 + (x_2 - p)^2$. 
Finally, the desired position $D(x_3, y_3)$ of the particle is given by 
\begin{subequations}
\begin{alignat}{2}
x_3 & = p \pm \,l \sqrt{ \frac{\sin^2\theta_i + e^2 \cos^2 \theta_i } {1 + m_3^2}},\\
y_3 & = q + m_3(x_3 - p),
\end{alignat}
\label{eq:fxy}
\end{subequations} 
where $m_3 = \tan (\theta_r + \tan^{-1} (m_2))$ is the slope of line $BD$.
In general, depending on the net force acting on the particle, the particle may reflect 
multiple times at the channel boundaries to reach the final position. 

The Langevin equation $(\ref{Langevin2})$ is solved by using the standard stochastic Euler algorithm over $2 \times 10^4$ trajectories with the reflection boundary conditions. 
Numerically, the nonlinear mobility and the effective diffusion coefficient are 
calculated as
\begin{align}
 \mu(f) &:= \lim_{t\to\infty} \frac{\langle x(t) \rangle}{ t \, f},\\
 D_{eff}  &:= \lim_{t\to\infty} \frac{\langle x^2(t) \rangle - \langle x(t) \rangle ^2 }{2 \,t}.
 \end{align}

\section{Diffusion of noninteracting particles} \label{sec:3}

The diffusion of noninteracting particles in a symmetric channel has been studied 
earlier both analytically and numerically by H\"anggi and co-workers.  \cite{Burada_prl,Burada_phd,Marten,Burada_pre} 
Under the assumption of a fast equilibration in the transverse direction of the channel, the dynamics of the system can be approximately described by means of a one-dimensional kinetic equation. The latter is obtained after integrating out the 
transverse coordinate ($y$) from the two-dimensional Smoluchowski equation.
The resultant one-dimensional probability density $P(x,t)$, which 
obeys the Fick-Jacobs equation, reads \cite{Burada_prl,Burada_pre}
\begin{equation}
\frac{\partial P(x, t)}{\partial t} = \frac{ \partial}{\partial x} D(x) \left\{\frac{ \partial P(x, t)}{\partial x} + A'(x) P(x, t)\right\}, 
\end{equation}
where $D(x) = 1/\left( 1 +  \omega' (x)^2 \right)^{1/3}$ is the position dependent 
diffusion coefficient for a 2D system, \cite{DR_pre} 
the dimensionless free energy is given by $A(x) =  -f x - \ln (2 \omega (x))$, and $2\omega (x)$ is the local width of the two-dimensional channel.
Remarkably, the free energy assumes the form of a periodic tilted 
potential whose barrier height is a function of the temperature. \cite{Burada_prl}
   
The corresponding transport characteristics, the nonlinear mobility and the effective diffusion coefficient, of the Brownian particles can be calculated using the mean first passage time (MFPT) approach \cite{Burada_prl} 
and are given by 
\begin{subequations}
\begin{equation}\label{mob}
 \mu(f) := \frac{\langle \dot{x} \rangle }{f} 
   = \frac{1 - e^{-f}}{f \displaystyle{\int_{0}^{1} I(z) \, dz}}, 
\end{equation}
 \begin{equation}\label{dif}
 \frac{D_{eff}}{D_0}  = \frac{\displaystyle \int_{0}^{1} \int_{x-1}^{x} \frac{D(z)}{D(x)}\, \frac{e^{A(x)}}{e^{A(z)}} \, [I(z)]^2   \, dx \, dz}{\displaystyle \left[\int_{0}^{1}  I(z) \, dz \right]^3},
 \end{equation}
\end{subequations}
where the integral $I(z) = e^{A(z)}/D(z) \int_{z-1}^{z}  e^{-A(y)} \, dy $.
    
\begin{figure}[t]
\centering
\includegraphics[scale = 1.0]{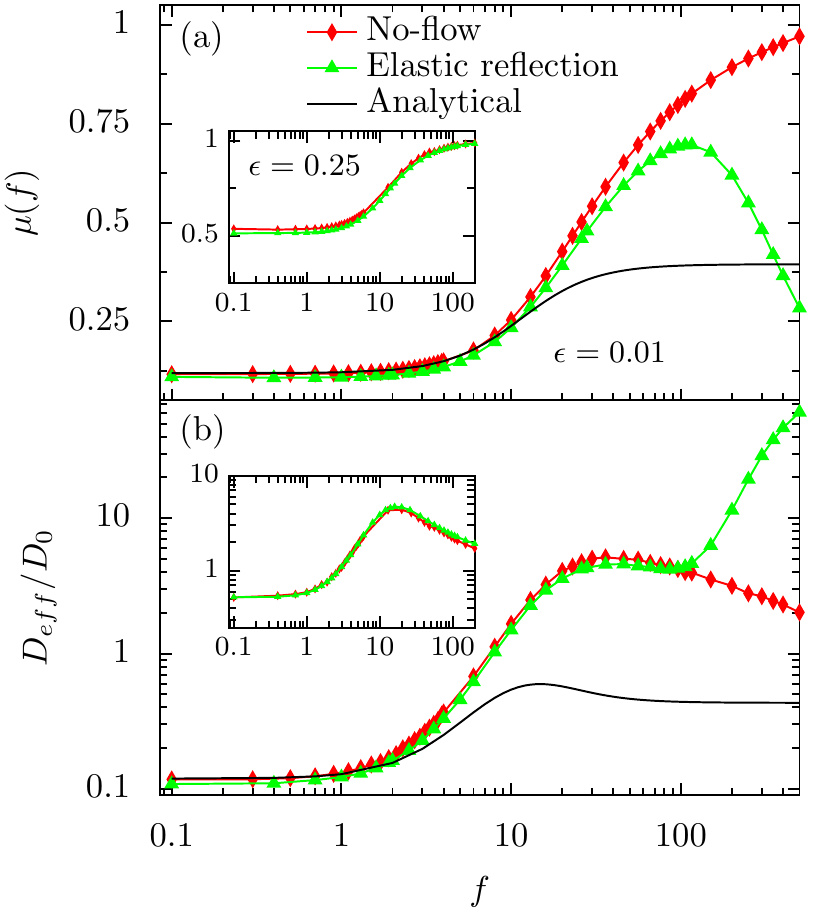}
\caption{ (Color online) Numerically simulated and analytically calculated, the nonlinear mobility $\mu(f)$ as a function of the scaling parameter $f$ is depicted in 
(a) for the symmetric channel with the small aspect ratio $\epsilon = 0.01$. 
The corresponding scaled effective diffusion coefficient $D_{eff}/D_0$ is depicted in (b). The insets depict the numerically simulated $ \mu(f)$ and $D_{eff}/D_0$  as a function of $f$ for the symmetric channel with a moderate aspect ratio $\epsilon = 0.25$. 
The other parameters of the channel are $L = 1$ and $\omega_\mathrm{max} = 2.02$.}
\label{fig:graph1}
\end{figure} 

Figure~\ref{fig:graph1} depicts the dependence of the nonlinear mobility and the effective diffusion coefficient on the scaling parameter $f$ with both no-flow and elastic reflection boundary conditions for two different symmetric channels with small and moderate aspect ratios $\epsilon = 0.01 \,\text{and}\,  0.25$, respectively. 
Note that in this case, the interaction between the particles is ignored such that the calculated quantities can be compared with the earlier predictions. \cite{Burada_prl,Burada_pre,Burada_cpc} 
Figure~\ref{fig:graph1} shows that, for $\epsilon = 0.01$, there is a clear difference between the results obtained using no-flow and elastic reflection boundary conditions at higher scaling parameter values. In the latter case, the channel boundaries exert an equal and opposite force on the incident particle. Thus, the particle takes multiple reflections within a cell before proceeding to the next one. 
However, the net motion is always in the positive $x-$ direction as the external force is acting along that direction. 
As a result, the average survival time $ \tau(f)$ of the particles \cite{Igor,Macro} increases in a cell.
This leads to a decrease in the mobility and an increase in the effective diffusion compared to the no-flow case. 
Whereas, for a moderate confined geometry, i.e., $\epsilon = 0.25$, these two boundary conditions lead to the same results (see the insets of Fig.~\ref{fig:graph1}).
Because, for this case, the average survival time is less compared to that of the previous case. 
Note that the considered channel is highly confined ($L = 1$), and as expected, the analytical findings agree well with the numerical results only in the small $f$ 
regime. \cite{Burada_pre}

\begin{figure}[t]
\includegraphics[width=0.55\linewidth]{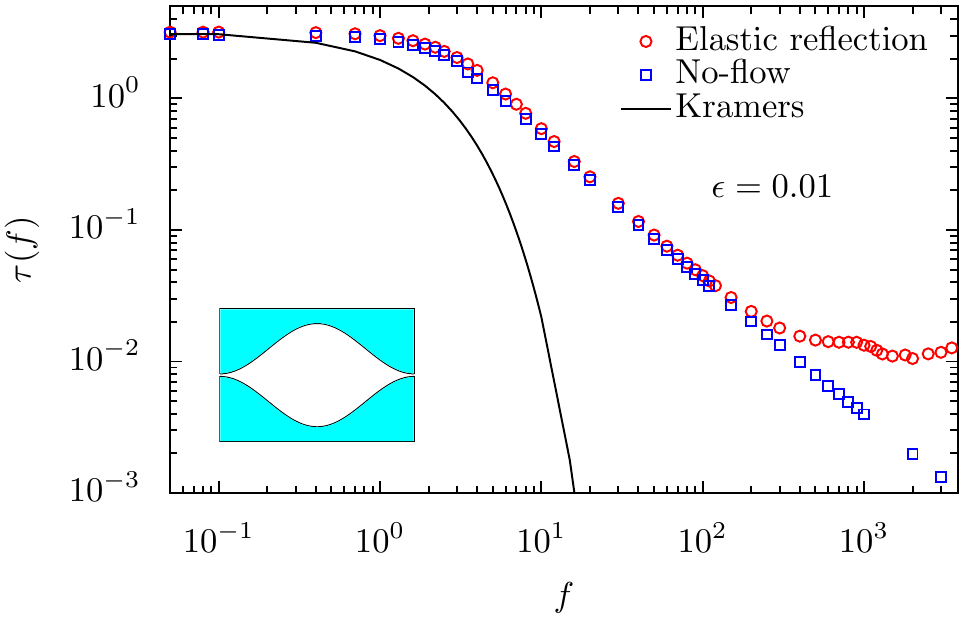}
\caption{(Color online) The average survival time $ \tau(f)$ as a function of the scaling parameter $f$. 
The solid line corresponds to the inverse of the Kramers-Smoluchowski rate calculated using the effective potential $A(x)$ (see the text).
The symbols correspond to the numerical results obtained for a single cell of the channel.
The other parameters of the channel are $L = 1$ and $\omega_\mathrm{max} = 2.02$.}
\label{fig:survival}
\end{figure}

To understand the average survival time of the Brownian particles in a given cell, we have calculated the Kramers-Smoluchowski rate \cite{Kramers,Burada_epjb} using the effective potential $A(x)$, 
which, in dimensionless units, is given by 
$ r_k(f) = \left(\sqrt{A^{''}(x_\mathrm{min}) | A^{''}(x_\mathrm{max})| }/(2 \pi)\right) e^{-\Delta A(f)} $,
where the primes refer to the double derivative with respect to $x$, $x_\mathrm{min}$ and $x_\mathrm{max}$ denote the positions of the minimum and maximum of the effective potential, respectively.
Therefore, the average survival time calculated using the Kramers-Smoluchowski rate is given by $ \tau(f)$ = $1/r_k (f)$.

Figure~\ref{fig:survival} depicts the dependence of the average survival time $\tau(f)$ on the scaling parameter $f$ in a symmetric channel consists of only one cell with a small aspect ratio $\epsilon = 0.01$ and the periodicity $L = 1$. 
As the force is acting along the channel direction, the average survival time decreases with increasing the scaling parameter $f$.
Note that as expected, the Kramers-Smoluchowski rate agrees well with the numerical simulations only in the limit $f \rightarrow 0$. 
Compared to the no-flow boundary conditions, $\tau(f)$ calculated using the 
elastic reflection boundary conditions decreases only up to an optimal value of the scaling parameter $f$. On further increasing $f$, $\tau(f)$ starts to deviate from the monotonically decreasing response due to multiple reflections that particles make at the channel boundaries. 
Thus, the average survival time of the particles in a given cell increases. 
As a result, the nonlinear mobility decreases and the effective diffusion coefficient increases for higher $f$ values.

\begin{figure}[htb]
\includegraphics[width=.5\linewidth]{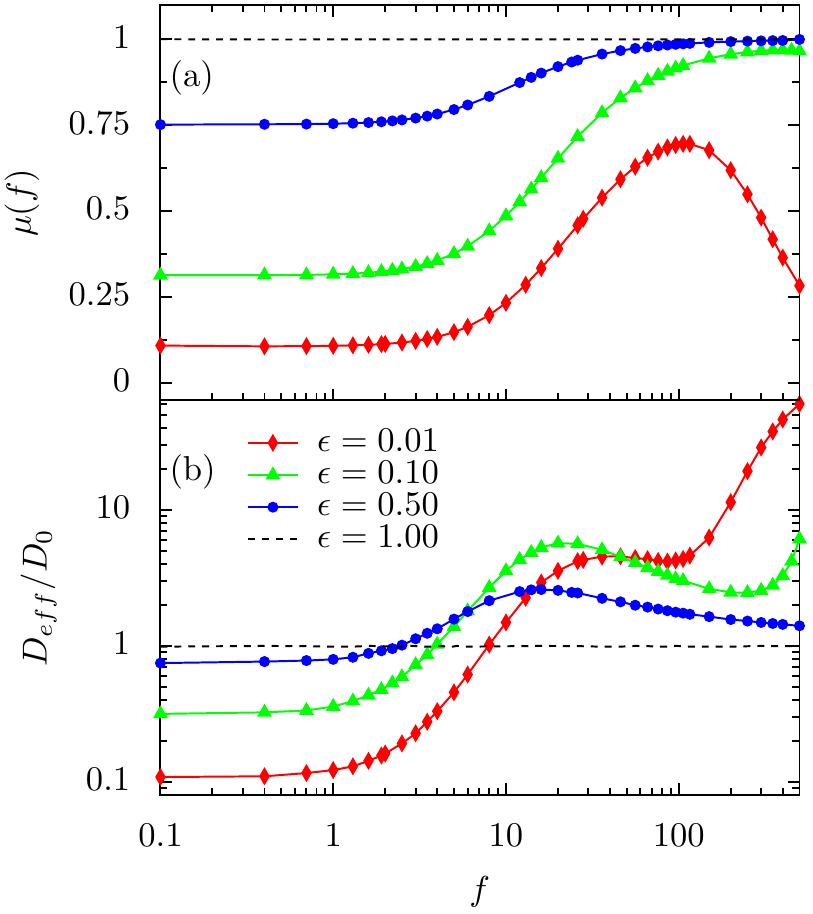}
\caption{(Color online) Numerically simulated, the nonlinear mobility $ \mu(f)$ as a function of the scaling parameter $f$ is depicted in 
(a) for different channel aspect ratios.  
The corresponding scaled effective diffusion coefficient $D_{eff}/D_0$ 
is depicted in (b).
The other parameters of the channel are $L = 1$ and $\omega_\mathrm{max} = 2.02$.}
\label{fig:graph2}
\end{figure}
 
Figure~\ref{fig:graph2} shows the behavior of the nonlinear mobility and the effective diffusion coefficient as a function of the scaling parameter $f$ for a symmetric channel with different aspect ratios. Note that the transport properties are greatly influenced by the channel geometry. 
For a small aspect ratio, the nonlinear mobility exhibits a nonmonotonic behavior and the effective diffusion is greatly enhanced. 
As mentioned earlier, this is due to the elastic reflection boundary conditions for which the average survival time of the particles increases at higher $f$ values.
However, as the aspect ratio of the channel increases, the average survival time decreases. This leads to a higher mobility and a lower effective diffusion. 
As one would expect, for $\epsilon = 1$, i.e., for a flat channel, both 
the nonlinear mobility and the effective diffusion coefficient equal to the bulk values 
for all $f$ (see Fig.~\ref{fig:graph2}).
Note that the analytical results are valid for smaller $\epsilon$ values only. \cite{Burada_phd}

\begin{figure}[htb]
\includegraphics[width=1.0\linewidth]{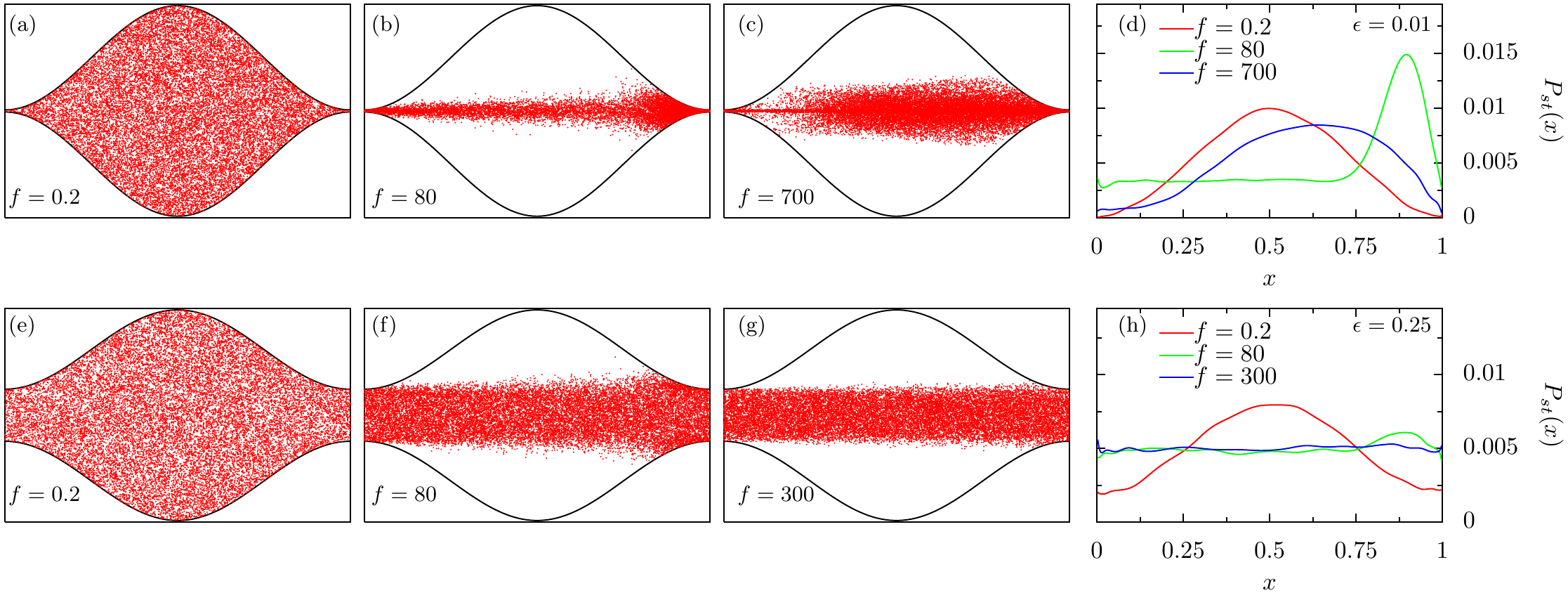}
\caption{(Color online) Steady state distribution of noninteracting Brownian particles, mapped into a single cell, for various values of the scaling parameter $f$. 
For the top panel, (a)-(c), $\epsilon = 0.01$ and for the bottom one, (e)-(g), 
$\epsilon = 0.25$. 
The corresponding normalized probability distributions are depicted in (d) and (h).
The other parameters of the channel are $L = 1$ and $\omega_\mathrm{max} = 2.02$.}
\label{fig:Pdist1}
\end{figure}

Figure~\ref{fig:Pdist1} depicts the steady state distribution and the corresponding normalized probability distribution of noninteracting particles for various values of the scaling parameter $f$. 
Note that since the channel is periodic, we have mapped the positions of the particles into a single cell.
Here, we have chosen two different channels, one with a small aspect ratio and the other with a moderate aspect ratio. 
For small $f$ values, particles distribute uniformly along the transverse direction inside a cell [see Figs. \ref{fig:Pdist1}(a) and \ref{fig:Pdist1}(e)], satisfying the equilibration assumption. \cite{Jacobs,Zwanzig,DR_pre,Kali1_pre} 
The corresponding normalized probability distribution $P_{st}(x)$ in the channel direction is Gaussian. 
For $\epsilon = 0.01$, at $f = 80$, particles tend to focus in the middle and the exit of the channel evidencing the failure of the equilibration assumption. \cite{Burada_phd,Burada_pre}
This is reflected in the corresponding normalized probability distribution [see Fig.~\ref{fig:Pdist1}(d)].
This is because the work done on the particles is greater than the thermal 
noise present in the system. 
By further increasing $f$, the effect of reflection boundary conditions starts to appear. The average survival time of the particles increases. 
This is reflected in an enhanced distribution of particles in the transverse direction 
[see Fig.~\ref{fig:Pdist1}(c)] and the corresponding probability distribution along the channel direction [see Fig.~\ref{fig:Pdist1}(d)].
However, for the channel with a moderate aspect ratio, the effect of reflection boundary conditions disappears. 
At higher force values, multiple reflections of the particles at the channel boundaries decrease because the bottleneck width of the channel is high. In other words, the average survival time of the particles in a given cell decreases and particles move in the middle of the channel [see Fig.~\ref{fig:Pdist1}(g)]. As a result, the corresponding normalized probability distribution is flat [see Fig.~\ref{fig:Pdist1}(h)].

\section{Inelastic reflection boundary conditions} \label{sec:4}

In this section, we focus on the effect of inelastic reflection boundary conditions on the transport properties of noninteracting Brownian particles. 
Here, we restrict to the channel with a small aspect ratio $\epsilon = 0.01$.  
Note that the strength of inelastic reflection is characterized by the coefficient of restitution $e$, which varies between $[0,1]$. 
$e = 1$ means a perfectly elastic reflection in which there is no loss of energy of the particle during reflection at the channel boundary and 
$0 < e < 1$ means an inelastic reflection in which there is a loss of energy. 
$e = 0$ means a perfectly inelastic reflection in which the kinetic energy of the particle along the normal to the channel boundary is zero. 
In a way, it is similar to the sticky boundary conditions used in the earlier studies.\cite{Bauer}
Note that the momentum of a particle is conserved for both the elastic and the inelastic reflection. 
However, the energy of a particle is not conserved for the inelastic reflection.

\begin{figure}[htb]
\includegraphics[width=.5\linewidth]{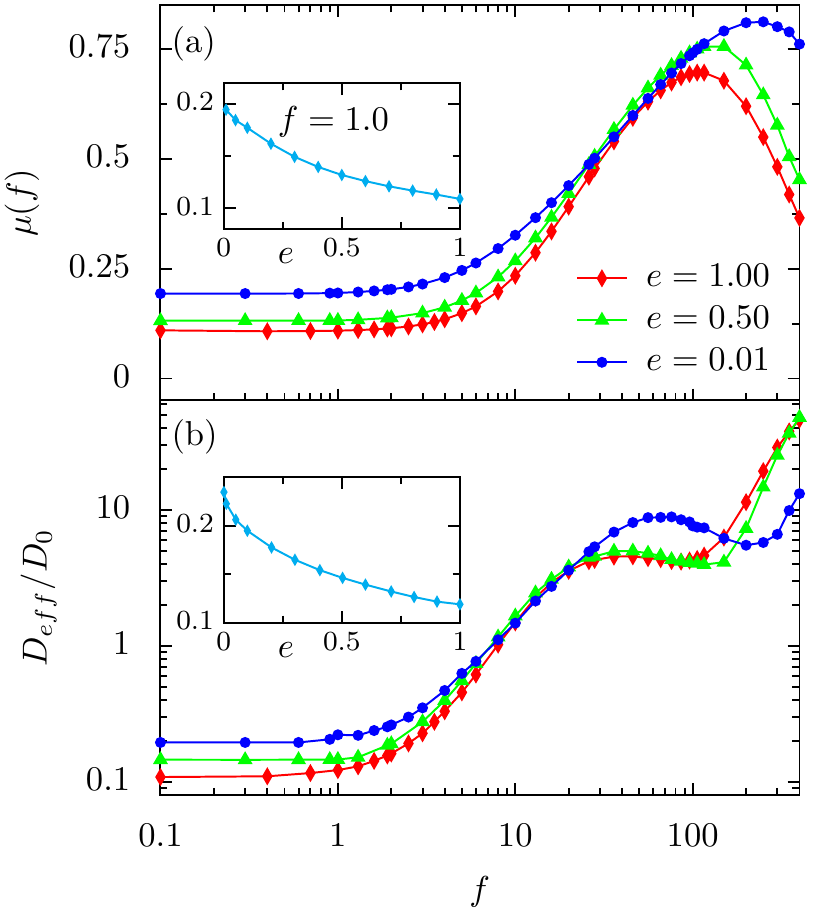}
\caption{(Color online) Numerically simulated, the nonlinear mobility $ \mu(f)$ as a function of the scaling parameter $f$ for various values of the coefficient of restitution $e$ is depicted in (a). 
The corresponding scaled effective diffusion coefficient $D_{eff}/D_0$ is depicted in (b). 
The insets depict $ \mu(f)$ and $D_{eff}/D_0$ as a function of $e$ for $f = 1$. 
The other parameters of the channel are $L = 1$ and $\omega_\mathrm{max} = 2.02$.}
\label{fig:Inelastic}
\end{figure}

Figure~\ref{fig:Inelastic} depicts the behavior of the nonlinear mobility and the effective diffusion coefficient as a function of the scaling parameter $f$ for various values of the coefficient of restitution $e$.   
Note that inelastic reflection boundary conditions can cause two effects to the particle: 
$(i)$ reduce its energy and 
$(ii)$ force to move along the channel boundary after the reflection.
In the limit $e \rightarrow 0$, the particle slows down quickly after the reflection due to a considerable amount of energy loss and moves along the channel boundary.
As a result, for small $f$ values, both the nonlinear mobility and the effective diffusion coefficient slightly increase on decreasing 
$e$ (see insets of Fig.~\ref{fig:Inelastic}). 
However, for a fixed $e$, as the $f$ increases, both the mobility and effective diffusion increase. 
At higher $f$, due to reflection boundary conditions, mobility decreases, and effective diffusion is greatly enhanced. 
Interestingly, in a narrow regime of moderate $f$ values, both the mobility and the effective diffusion 
are unaffected by $e$.     

\section{Diffusion of interacting particles} \label{sec:5}

\begin{figure}[thb]
\includegraphics[width=.5\linewidth]{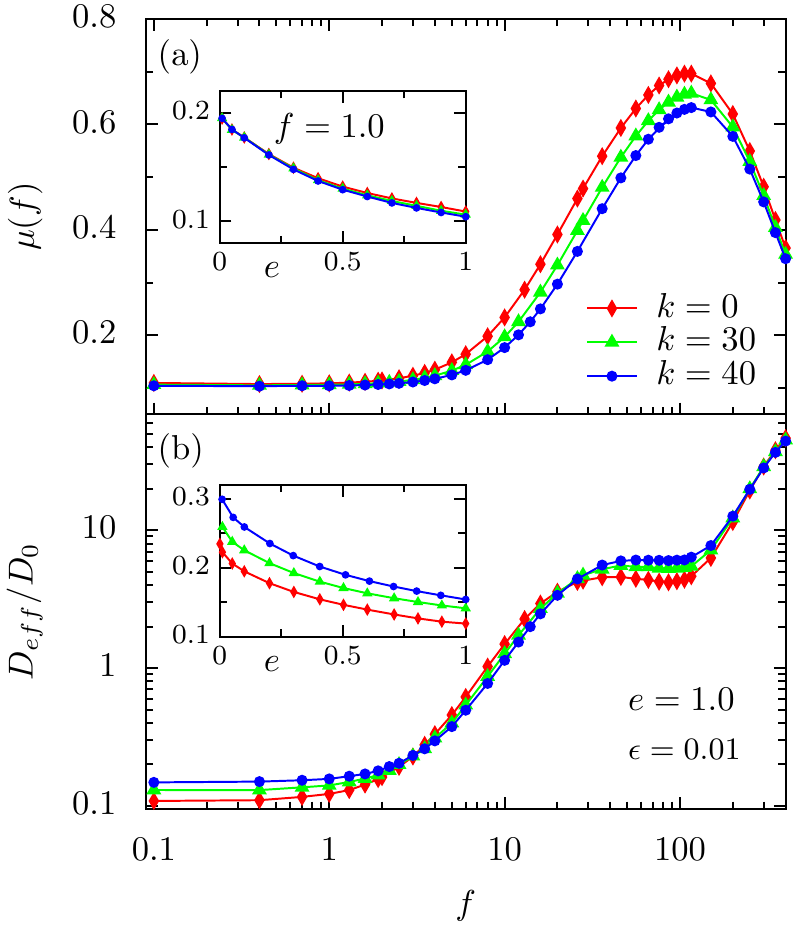}
\caption{(Color online) Numerically simulated, the nonlinear mobility $ \mu(f)$ as a function of the scaling parameter $f$ for various strengths of the scaled interaction $k$ is depicted in (a).
Here, we have fixed the coefficient of restitution $e = 1$ and the aspect ratio of the channel $\epsilon = 0.01$.
The corresponding scaled effective diffusion coefficient $D_{eff}/D_0$ is depicted in (b). The insets depict $\mu(f)$ and $D_{eff}/D_0$ as a function of $e$ for various strengths of the scaled interaction $k$ at $f = 1$.
The other parameters of the channel are $L = 1$ and $\omega_\mathrm{max} = 2.02$.}
\label{fig:Intgraph1}
\end{figure}

In this section, we study the transport of interacting Brownian particles in a channel 
with a small aspect ratio ($\epsilon = 0.01$). 
In the limit of high density of particles, the short range interaction force between the particles is similar to the Vicsek interaction \cite{Vicsek} as we have discussed in Sec.~\ref{sec:model}.  
Note that the approximate 1D analytical description cannot be used for this case. 
Therefore, we stick to the numerical simulations with 
the reflection boundary conditions. 
Figure~\ref{fig:Intgraph1} depicts the behavior of the nonlinear mobility and the effective diffusion coefficient as a function of the scaling parameter $f$ for different scaled interaction strengths $k$.
Note that the short range interaction force between the particles, for the optimal values of $f$, can cause two effects: 
$(i)$ reduce the biased motion of the particles on increasing $k$, resulting in a decrease of the nonlinear mobility and 
$(ii)$ particles tend to move in random orientations on increasing $k$, resulting 
in an increase of the effective diffusion coefficient.
However, the qualitative behavior of the transport characteristics is unchanged with the scaled interaction strength. 
Also, in the limit $f \to 0$, as expected, particles interaction does not contribute to the average mobility but influences the effective diffusion (see Fig.~\ref{fig:Intgraph1}).
In the other limit, i.e., $f \to \infty$, the scaled applied force dominates over the scaled interaction force. 
As a result, the nonlinear mobility and the effective diffusion coefficient are 
independent of the scaled interaction strength.
However, qualitatively, the impact of the coefficient of restitution $e$ on these quantities remains similar to that observed in Fig.~\ref{fig:Inelastic} 
(see insets of Fig.~\ref{fig:Intgraph1}).

\section{Conclusions} \label{Conclusions}
      
With this work, we have shown that the transport of Brownian particles in symmetric channels with reflection boundary conditions exhibits distinct properties which have not been observed with no-flow boundary conditions. 
Using numerical simulations, we have investigated that the transport properties can be effectively controlled by the scaling parameter $f$, the aspect ratio of the symmetric channel $\epsilon$, the coefficient of restitution $e$, and the scaled interaction strength $k$.
We have observed that for the channel with a small aspect ratio, neither the nonlinear mobility nor the effective diffusion coefficient approaches to the bulk values in the limit $f \to \infty$. 
The nonlinear mobility exhibits a nonmonotonic behavior, and the effective diffusion shows a rapidly increasing behavior. It is a clear signature of the reflection boundary conditions.  
Also, we have numerically demonstrated that both the nonlinear mobility and the effective diffusion coefficient can be enhanced with inelastic reflection boundary conditions.
In addition, for the optimal values of $f$, the interaction between the particles leads to a decrease in the mobility and an increase in the effective diffusion. 
However, in the $f \to \infty$ limit, the interaction between the particles does not influence the transport characteristics.
These results evidence the usefulness of reflection boundary conditions which may occur naturally in systems such as nanoporous materials, zeolites, microfluidic devices, ratchets, artificial channels, etc. \cite{Karger,Hille, Reza, Han,Brangwynne,Freyhardt,Davis_nature1,Marchesoni_rev, Burada_prl2} 
Furthermore, the presented model could be used to design artificial channels for 
controlled drug release, \cite{Siegel} 
transport of particles, \cite{Siwy,Chou,Berezhkovskii} and
particle separation. \cite{Burada_prl2,Keller,IDeyenyi,Li_pre}

\section{ACKNOWLEDGMENTS}

This work was supported by the Indian Institute of Technology Kharagpur under the Grant No. IIT/SRIC/PHY/TAB/2015-16/114.

\end{document}